\documentclass[english,smallextended]{article}       

\usepackage[pdftex]{graphicx}
\usepackage{amssymb}
\usepackage{amsmath}
\usepackage{color}
\usepackage{authblk}

\usepackage{mciteplus}
\usepackage[hidelinks,bookmarksopen]{hyperref}
\usepackage{epstopdf}
\begin{document}

\title{Is all-electrical silicon quantum computing feasible in the long term?}
\author{Elena Ferraro}
\affil{CNR-IMM Agrate Unit, Via C. Olivetti 2, 20864 Agrate Brianza (MB), Italy\\
elena.ferraro@mdm.imm.cnr.it}

\author{Enrico Prati}
\affil{CNR-IFN, Piazza Leonardo da Vinci 32, 20133 Milano, Italy}

\maketitle
\begin{abstract}
The development of the first generation of commercial quantum computers is based on superconductive qubits and trapped ions respectively. Other technologies such as semiconductor quantum dots, neutral ions and photons could in principle provide an alternative to achieve comparable results in the medium term. It is relevant to evaluate if one or more of them is potentially more effective to address scalability to millions of qubits in the long term, in view of creating a universal quantum computer. We review an all-electrical silicon spin qubit, that is the double quantum dot hybrid qubit, a quantum technology which relies on both solid theoretical grounding on one side, and massive fabrication technology of nanometric scale devices by the existing silicon supply chain on the other.      
\end{abstract}


\section{Introduction}
The assessment of a quantum computer requires a solid and versatile platform in terms of fabrication, scalability, integration and reliability. Today's quantum processing units (QPUs) are implemented by a variety of physical systems competing to achieve the goal of ultimate universal quantum computing. Key ingredients are both high quality physical qubits and fault tolerance by quantum error correction. Semiconductor qubits encode quantum information by the spin of either electrons or holes confined through artificial atoms such as quantum dots and individual donor atoms. Spin qubits have already proved feasibility \cite{crippa2019gate} and integration. They are in principle suitable for mass scalability \cite{jehl2011mass} and sufficient reliability \cite{Kim-2015}. The fabrication techniques developed by the current microelectronics industry allow in principle relevant advantages in terms of scalability, integrating multiple qubits in micrometric scale circuits \cite{Rotta-2016,Rotta-2017}. Spin qubits are being developed by academia \cite{Morton-2011,Veldhorst-2017}, large pre-industrial fabrication facilities (LETI, IMEC) \cite{maurand2016cmos,IMEC}, and a large company (Intel).
Although the integration of multiple silicon spin qubits with Complementary Metal–Oxide–Semiconductor (CMOS) control electronics is not straightforward, significant steps forward have been done very recently binding the assessment of a full CMOS approach \cite{maurand2016cmos} that combines classical electronics with quantum circuits on the same substrate operating at cryogenic temperatures \cite{tagliaferri2016modular,Homulle-2017}. 

A major obstacle to the implementation of fault-tolerant quantum computers at large scale is indeed represented by the intrinsic fragility of  quantum  states. Qubits are inevitably coupled to the environmental degrees of freedom causing a loss of coherence that affects their operations. In superconductive qubits, the mitigation is achieved by the combination of millikelvin temperature and the relative large size of the superconductive circuit of the order of tens of microns, which make the device insensitive to nanometric scale variability. Trapped ions have the advantage of being potentially worked out at room temperature thanks to the stability of the atomic states suspended in a high vacuum, and a manipulation at millimetric length scale. Spin qubits in silicon are deeply affected by the influence of magnetic noise, i.e. the dephasing due to the presence of nuclear spins, as well as by the influence of charge noise, arising from the fluctuations of the energy levels on each quantum dot due to noisy gate voltages or the environment \cite{prati2007microwave,prati2008effect}. The use of purified isotopes with zero nuclear spin ($^{28}$Si) is a valuable strategy to reduce significantly magnetic noise due to the hyperfine interaction \cite{prati2013quantum}, in addition several techniques have been discussed which partly decouple the qubit from magnetic noise \cite{West-2012}. To overcome the issue related to the ubiquitous electrical noise it is shown that it is favorable to operate the qubit on the so-called ‘sweet spots’ less susceptible to noise \cite{Shim-2016}, providing in this way a longer qubit lifetime. Moreover the analysis of the qubit decoherence must also include the electron–phonon interaction. It can arise from an inhomogeneous deformation of the crystal potential, resulting in an alteration of the band-gap (in all semiconductors), and from a homogeneous strain due to piezo-electric effect (in crystals without structure inversion symmetry, i.e. GaAs, not in Si). This contribution, contrary to the hyperfine interaction that is reduced with a large Zeeman splitting, is instead enhanced by an applied magnetic field. Finally, in silicon also the physics of the conduction band minima could affect the dynamics. In bulk silicon the electrons lying in the minimum of the conductance band carry sixfold valley degeneracy \cite{de2012geometrical}. Confinement along one direction lifts valley degeneracy. Only if the qubit energy splitting is smaller with respect to the valley splittings the latter contribution can be disregarded. The degeneracy of the maxima of valence band at k=0 play the analogous role for hole-spin qubits, that are operated exploiting the spin-orbit coupling.

Lastly, fast gate operations must follow the adiabaticity requirement. A quantum gate is adiabatic when the parameters of a qubit prepared in an arbitrary superposition of eigenstates of the Hamiltonian change slowly during the time so that the transitions between eigenspaces is negligible. On one hand the pulse sequences used in adiabatic protocols are more resilient against pulse errors but on the other, being slower, they could be more sensitive to charge noise. For this reason, it is relevant to connect the fidelity of quantum gates with the charge noise. 

In this perspective article we focus on a all-electrical realization of quantum dot qubit in silicon, i.e. the three electrons in a double quantum dot hybrid qubit. It represents a compromise between a charge qubit operating when the detuning is small and close to zero, and a spin qubit for large detuning values. As said, it is realized electrostatically confining three electrons in a double quantum dot and the logical states are defined in the spin subspace with total spin $S=1/2$ and vertical component $S_z=-1/2$. 

The article is organized as follows: Section 2 discusses the silicon qubit and the theoretical model of the hybrid all-electrical qubit, Sections 3 the theory involving quantum information processing by hybrid qubits, Section 4 the experiments carried at cryogenic temperature, while Section 5 discusses the scaling perspectives.

\section{Qubit based on silicon quantum dots}
How to realize a spin-based quantum computer is still the subject of a heated debate. Several reliable qubit implementations are based on both GaAs III–V compound  \cite{Kikkawa-1998,Koppens-2008,Gamble-PRB2012} and group IV semiconductors such as Si  \cite{Zwanenburg-2013,Morello-2010,Pla-2012,Pla-2013,Veldhorst-2014,Veldhorst-2015} and SiGe \cite{maune2012coherent} compound. Electrically gated quantum dots in silicon, combined with integrated classical electronics  \cite{Russ-2017,Kloeffel-2013,Hanson-2007}, assure excellent manipulation and potentially scalability. Continuous progresses have been made over the past decade for single and two qubits operations in single \cite{Yoneda-2018,Zajac-2018,Loss-1998,Morton-2011,Veldhorst-2014,Kawakami-2014}, double \cite{Nichol-2017,Petta-2005,Maune-2012,XianWu-2014,Barnes-2016,Taylor-2007}, and triple quantum dots \cite{DiVincenzo-2000,Nakajima-2016,Ludwig-2007,Granger-2010} respectively. However, in order to be competitive with respect to the current alternative technologies such as superconductors and trapped ions, it is necessary to improve the qubit performance in terms of the number of gate operations that can be executed within the coherence time of the spin qubit. 

Table \ref{comparison1} contains an overview of the more appealing spin qubits in silicon (quantum dots and donors), reporting the main figures of merit: the qubit frequency, the coherence time (Ramsey experiment), the dephasing time (spin echo) and the quality factor related to a $\pi$-pulse operation.

\begin{center} 
\begin{table}[htbp!]
\begin{tabular}{ |p{2.6cm}|p{1.6cm}|p{1.6cm}|p{1.8cm}|p{1.6cm}|p{1.8cm}|p{0.5cm}|}
\hline
Qubit & Material & f(MHz) & $T_2^{\ast}(ns)$ & $T_2(ns)$ & $Q\equiv T_2^{\ast}/T_{\pi}$ & Ref.\\
\hline
Single spin & Si/SiGe & $\sim 5$ & $\sim 9\times 10^2$ & $3.7\times 10^4$ & $\sim 9$ & \cite{Kawakami-2014}\\ \hline
Single spin & $^{28}$Si & $\sim 0.3$ & $\le 1.2\times 10^5$ & $1.2\times 10^6$ & $\le 80$ & \cite{Veldhorst-2015}\\ \hline
Donor spin $(e^-)$ & P in $^{nat}$Si & $\sim 3$ & $55$ & $2\times 10^5$ & $\le 1$ & \cite{Pla-2012}\\ \hline
Donor spin $(e^-)$ & P in $^{28}$Si & $\sim 0.2$ & $\sim 3\times 10^5$ & $1\times 10^6$ & $\sim 108$ & \cite{Muhonen-2014}\\ \hline
Singlet-Triplet & Si/SiGe & $\sim 351$ & $>1\times10^3$ & n.a. & n.a. & \cite{Takeda-2019arxiv} \\ \hline
Hybrid & Si/SiGe & $\sim 1\times 10^{4}$ & $\sim 11$ & $\sim 40$ & $\sim 250$ & \cite{Kim-2014}\\\hline
\end{tabular}
\caption{Comparison of different spin-based qubits.}\label{comparison1}
\end{table}
\end{center}

Starting from these considerations, hybrid qubits based on three electrons in a double quantum dot provides a balanced compromise among fabrication, tunability, fast gate operations, manipulation and scalability \cite{Shi-2012}.

\paragraph{Hybrid qubit} 
One of the most appealing property is related to the manipulation, that is purely electrical when gate operations are performed. This feature enables much faster gate operations than using ac magnetic fields, inhomogeneous dc magnetic fields or mechanisms based on spin–orbit coupling \cite{Nadj-2010,Jock-2018,Corna-2018}. The use of oscillating magnetic or electrical fields or quasi-static Zeeman field gradient, which is mandatory in singlet-triplet qubits, is here unnecessary. The knobs that suffice for all the gate operations are the exchange interactions between pairs of spins. Moreover such an implementation presents a clear analogy with the qubit proposed in Ref. \cite{DiVincenzo-2000} with three electrons operating in a triple quantum dot. 

The basis state of the Hilbert space containing three electron spins, representing one qubit, is written in the computational basis via Clebsch–Gordan coefficients \cite{Ferraro-2015-qip}. To encode the hybrid qubit, the two-dimensional subspace of three-spin states with quantum numbers $S=1/2$ and $S_z=-1/2$ is used. The logical basis is constituted by singlet and triplet states of a pair of electrons, for example the pair in the left dot, combined with the single spin state of the third electron, localized in the right dot, forming two virtual qubit states
\begin{equation}\label{01}
|0\rangle_L\equiv|S\rangle|\!\!\downarrow\rangle, \qquad |1\rangle_L\equiv\sqrt{\frac{1}{3}}|T_0\rangle|\!\!\downarrow\rangle-\sqrt{\frac{2}{3}}|T_-\rangle|\!\!\uparrow\rangle
\end{equation}
where $|S\rangle$, $|T_0\rangle$ and $|T_-\rangle$ are respectively the singlet and triplet states $|S\rangle=\frac{|\!\uparrow\downarrow\rangle-|\!\downarrow\uparrow\rangle}{\sqrt{2}}$, $|T_0\rangle=\frac{|\!\uparrow\downarrow\rangle+|\!\downarrow\uparrow\rangle}{\sqrt{2}}$, $|T_-\rangle=|\!\!\downarrow\downarrow\rangle$ in the left dot, and $|\!\!\uparrow\rangle$ and $|\!\!\downarrow\rangle$ respectively denote a spin-up and spin-down electron in the right dot.

\begin{figure}[!htp]
\centering
\includegraphics[width=0.5\textwidth]{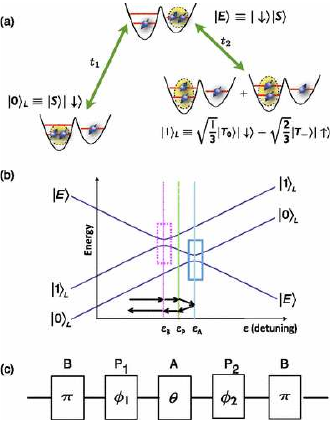}
\caption{(a) The hybrid qubit is a virtual qubit consisting of three physical spin qubits. The logical states are connected via the excited state $E$. (b) Energy diagram of the eigenstates of the Hamiltonian, showing energy range for control operations, as a function of the detuning $\epsilon$ between the two quantum dots that can be changed by varying the applied electrostatic potential in one of the two dots. (c) The circuit diagram for the gate sequence corresponding to the arrows in (b). The A and B gates correspond to charge qubit rotations, while P is a phase gate. Reprinted with permission from \cite{Koh-2012} Copyright 2012, American Physical Society http://dx.doi.org/10.1103/PhysRevLett.109.250503.}\label{1_Koh}
\end{figure} 

The schematic of the hybrid qubit is reported in Fig. \ref{1_Koh}a. In addition to the logical states, an (1,2) excited state $|E\rangle\equiv|\!\!\downarrow\rangle|S\rangle$ induces transitions between them through the tunneling amplitudes $t_1$ and $t_2$. The Hamiltonian model in the $\{|0\rangle_L, |1\rangle_L, |E\rangle\}$ basis is
\begin{equation}\label{H}
H=
\left( 
\begin{array}{ccc}
0 & 0 & t_1 \\
0 & E_{01} & -t_2\\
t_1 & -t_2 & \varepsilon
\end{array} 
\right),
\end{equation}
where $E_{01}$ represents the energy splitting between the
logical qubit states and $\varepsilon$ is the detuning between the two quantum dots that can be changed by varying the applied electrostatic potential in one of the two dot. The energies of $|0\rangle_L$, $|1\rangle_L$ and $|E\rangle_L$ as a function of the detuning $\varepsilon$ between the two dots are reported in \ref{1_Koh}b. The ground state changes charge occupation from (2,1) when $\varepsilon<\varepsilon_A$ to (1,2) when $\varepsilon>\varepsilon_A$. Two avoided crossing occur between $|0\rangle_L$ and $|E\rangle$ at $\varepsilon_A$ (blue box) and between $|1\rangle_L$ and $|E\rangle$ at $\varepsilon_B$ (dotted magenta box) through which pulse-gate transitions between $|0\rangle_L$ and $|1\rangle_L$ can be performed. Pulses of the detuning $\varepsilon_P$ are used to induce phase differences between the three states. An example of gating sequence to perform arbitrary rotations in the qubit space is indicated with arrows. Figure \ref{1_Koh}c shows the corresponding circuit diagram for the gate sequence.

\section{Theory}
This Section collects the main theoretical results on the hybrid qubit.

A compact theoretical description of the hybrid qubit through an effective model appeared in Ref. \cite{Ferraro-2014}. The general effective Hamiltonian is derived in terms of the spin operators of the three electrons and of their interactions, starting from the general Hubbard-like model. Technically, it is derived by defining a suitable projection operator according to the Schrieffer and Wolff method \cite{Schrieffer-1966}. The coupling constants, preserving an explicit dependence of all the parameters, are obtained analytically. The one qubit effective Hamiltonian is exploited to study two qubits systems. For a pair of interacting qubits there are two different connection designs, due to the asymmetry of the hybrid qubit. The Hamiltonian of both cases can be derived analytically as from \cite{Ferraro-2015-qip}. The evolution operator obtained from the one- and two- qubits effective Hamiltonian is a powerful tool that has been exploited in the derivation of a universal set of quantum gates, that is composed by single qubit operations for arbitrary rotations and the two-qubits controlled NOT (CNOT) gate \cite{Ferraro-2015-qip,DeMichielis-2015}. Single qubit gates for arbitrary rotations have been derived analytically by using the Euler angle method, if rotations by arbitrary angles about two orthogonal axes are available, i.e. $R_x(\phi)$ and $R_z(\theta)$. $R_x(\phi)$ ($R_z(\theta)$) is implemented by a two-step (three-step) sequence imposing that the evolution matrix of the entire sequence, that is the product between the unitary evolution matrix of each step, is equal to the corresponding rotation matrix $R_x(\phi)$ ($R_z(\theta)$) \cite{Ferraro-2018}. The sequences for the CNOT gates in the two different designs are obtained numerically with a search algorithm, that is a combination of a simplex-based and a genetic algorithms. At each iteration the sequences approaches the global minimum, featuring a reduced number of exchange steps and minimum interaction time. To exemplify a realistic condition, a double quantum dot in a Si nanowire is simulated by employing SDFT (spin density functional theory) predicting CNOT gate operations times in the range of 10 ns \cite{DeMichielis-2015}. 

Despite such pulses are not adiabatic, Ref. \cite{Koh-2012} reports a study on the implementation of fast pulsed gates that exploit the qubit level crossings. One- and two-qubits gates, X gate and CPHASE gate respectively, are implemented with dc quantum gates that are simpler with respect to the ac quantum gates. Such control sequences show fast gates in the subnanosecond regime with high fidelities. In Ref. \cite{Wong-2016}, Wong presents a theoretical study on the decoherence caused by 1/f charge noise and how to minimize the charge noise dependence in the qubit frequency. Optimal working points for ac gate operations that drive the detuning and tunnel coupling are also determined. The calculations show \textcolor{red}{X} gate fidelities up to 99.8\% that are exponentially sensitive to the 1/f detuning noise parameter, meaning that a small reduction in charge noise could improve significantly qubit fidelity.

How the hybrid qubit dynamics is affected by magnetic and electrical noise mainly due to fluctuations in the applied magnetic field and charge fluctuations on the electrostatic gates, is analyzed in \cite{Ferraro-QIP2018,Ferraro-2018-adv}. Magnetic noise originating from nuclear spins is negligible in natural Si and, even more, for isotopically purified Si. Fluctuations in the applied magnetic field and charge fluctuations on the electrostatic gates
adopted to confine the electrons, is taken into account including random magnetic field and coupling terms in the Hamiltonian deriving the behavior of the return probability as a function of time for initial conditions of interest. The evaluation of the coherence times, in correspondence of model parameters taking values of experimental interest, is done by extracting them through an envelope-fitting procedure on the return probabilities, giving a $T_2^*$ in the range from tens up to hundreds of ns, depending on the entity of the noise. 

The phonon-induced relaxation and decoherence processes is investigated by adopting the Bloch-Redfield theory in \cite{Ferraro-2019}. By employing a three-level effective model for the qubit and describing the environment bath as a series of harmonic oscillators in the thermal equilibrium states, the relaxation and decoherence times as a function of the bath spectral density and of the bath temperature are extracted obtaing results in the range of hundrens of ns. Moreover strongly driven silicon-quantum-dot hybrid qubit is studied in \cite{Yang-2017} and in \cite{Yang-2019PRA} in presence of 1/f charge noise. For X gate it is shown how the fidelity can  be improved reaching values $>99.9\%$, even in the presence of phonon dephasing.

Concerning the progress on the two-qubits case, a method for transferring single electrons by a pulsed gate is proposed in Ref. \cite{Mehl-2015-2}. The effect is achieved by applying fast voltage pulses at gates close to the quantum dots. The main challenge of the entangling gate is to avoid leakage to other states. Either a two-step procedure or an adiabatic manipulation protocol is proposed. The procedure provides a variety of leakage states and the couplings to them must be avoided during the operation. These couplings but also nuclear spin noise and charge noise introduce errors for the pulse-gated two-qubits operation that can not be overlooked. 

A powerful adiabatic entangling protocol based on capacitive couplings between hybrid qubit has been proposed \cite{Frees-2019}. Adiabatic protocols are more resilient against pulse errors than non-adiabatic pulses. They are less susceptible to leakage errors, but due to slower speeds they could be more to charge noise. For this reason it is crucial to study the effect of charge noise on the gate fidelities. The approach presented is based on the tuneable, electrostatic coupling between distinct charge configurations, and the new concept of a dynamical sweet spot (DSS) is developed. A simple pulse sequence that achieves an approximate DSS for a CZ gate is reported, with a significant improvement in the fidelity. The highest-fidelity sequence gives an average process fidelity of 99.08\%. A two-hybrid-qubits coupling scheme based instead on exchange interactions has also been proposed \cite{Yang-2019}. With this approach the qubits remain at their sweet spots during the whole operation. In the presence of realistic quasistatic and 1/f charge noise, the simulations for the CZ and ZCNOT gates show fidelities both $>99.9\%$.

Another fundamental aspect is related to the realization of devices able to interconnect remote sites composing the quantum circuit to transfer information \cite{Kandel-2019,Mills-2019}. In order to overcome the problem of interaction between distant qubits, different routes have been pursued: the SWAP chain protocol \cite{Rotta-2016,Rotta-2017} and the coherent tunneling by adiabatic passage (CTAP) scheme \cite{Ferraro-2015-prb,porotti2019coherent,porotti2019reinforcement}. The SWAP method is based on the sequential repetition of SWAP gates between adjacent qubits. The CTAP scheme consists of the tunneling of the three electrons localized initially in the first double quantum dot at the head of the chain to the end by adopting Gaussian pulses in a reversed sequence. The population transfer between two distant double quantum dots may be achieved without occupation in the internal quantum dots. 

\section{Experiments with spin qubits}
Experiments involving spin qubits in silicon have been reported by an increasing number of groups by a variety of different approaches. In order to be scalable, CMOS standard technology is likely to be preferred. Indeed the exploitation of the well assessed silicon industry could pave the way towards the large-scale quantum computation era based on the same silicon chip technology which played a prominent role in the current information age. To date, the only experiments reported on quantum dots based on a CMOS platform compatible with an industrial process are those based on the devices fabricated by LETI \cite{maurand2016cmos,crippa2019gate}, while INTEL is still working on its 300mm process line for qubit fabrication \cite{clarke2019spin}. In the following we focus on the literature of experiments of hybrid spin qubits in silicon, which could be fabricated with the same CMOS standard technology mentioned above.  

All the experimental results presented in this Section are based on devices realized is Si/SiGe heterostructures where electrostatically top-gated quantum dots are realized.

In 2012 Z. Shi and coworkers presented experimental evidence of the feasibility to implement hybrid qubit in silicon quantum dots \cite{Shi-2012}. They made the first experimental test on a Si/SiGe device and measured the triplet-singlet relaxation time in a single silicon dot, with $T_1=141\pm 12$ ms. They also demonstrated readout of the singlet and triplet states of two electrons in a silicon dot. They estimated theoretically dephasing times to be on the order of microseconds, a long time compared to quantum operations times.

In 2014 the fast hybrid qubit full control on the Bloch sphere was demonstrated \cite{Kim-2014}. A $\pi$-rotation times of less than 100 ps in two orthogonal directions has been demonstrated. Moreover high X and Z rotations fidelities are extracted finding respectively 85\% and 94\%.

Such results have been improved in \cite{Kim-2015} where quantum process tomography yields gate fidelity higher than 93\% (96\%) around the X (Z) axis of the Bloch sphere. By resonantly modulating the double dot energy detuning and employing electron tunnelling-based readout, fast Rabi oscillations and purely electrical manipulations of the three-electron spin states has been achieved. A Ramsey pulse sequence as well as microwave phase control are adopted to demonstrate universal single qubit gates. 

\begin{figure}[!htp]
\centering
\includegraphics[width=\textwidth]{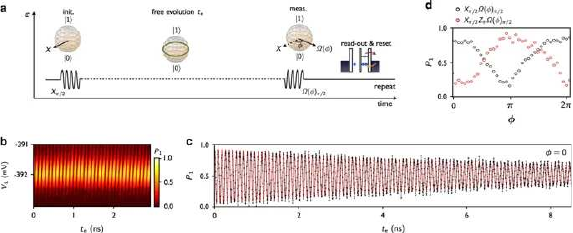}
\caption{(a) Schematic diagram of the pulse sequences to perform universal control of the qubit. (b) $P_1$ of the state to be $|1\rangle$ at the end of the driving sequence as a function of the voltage $V_L$ and the delay time $t_e$ for the initial condition $|Y\rangle=1/\sqrt{2}(|0\rangle+i|1\rangle)$. (c) $P_1$ as a function of $t_e$ with $V_L$=-391.7 mV, showing $\approx$11.52 GHz Ramsey fringes. (d) Effect of the phase $\phi$ of the second microwave pulse on the state $|Y\rangle$ (by applying $X_{\pi/2}$ on $|0\rangle$, black), and $|-Y\rangle$ (by applying $X_{\pi/2}$ and $Z_{\pi}$ on $|0\rangle$, red). Reprinted with permission from \cite{Kim-2015} Copyright 2015, Springer Nature.}\label{2_Kim}
\end{figure} 

In Fig. \ref{2_Kim}a the schematic diagram of the pulse sequences used to perform universal control of the qubit is reported, where $t_e$ is the delay time and $\phi$ is the phase of the second microwave pulse. The experimental measurements of Z axis rotation are reported in \ref{2_Kim}b: $P_1$ as a function of the voltage $V_L$ and $t_e$ for $|Y\rangle=1/\sqrt{2}(|0\rangle+i|1\rangle)$ and in \ref{2_Kim}c: $P_1$ as a function of $t_e$ with fixed $V_L=-391.7$ mV, showing $\approx$11.52 GHz Ramsey fringes. The red solid curve supplies the best fit parameter $T_2^\ast=11$ ns. The effect of the phase $\phi$ on the state $|Y\rangle$ applying two different pulses is shown in \ref{2_Kim}d. In both cases the oscillation of $P_1$ are visible. Moreover, they implement dynamic decoupling sequences on the hybrid qubit enabling coherence times longer than 150 ns.

\begin{figure}[!htp]
\centering
\includegraphics[width=\textwidth]{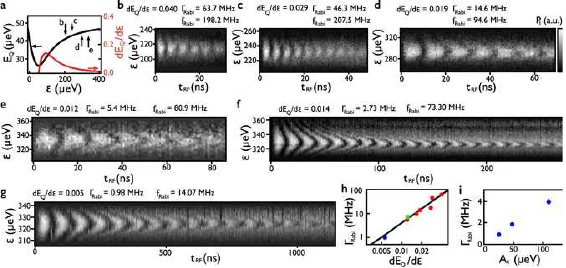}
\caption{(a) Plots of the qubit energy $E_Q$ (black) and its derivative (red) as a function of the detuning $\varepsilon$. (b)–(e) Rabi oscillations of the probability $P_1$ obtained at the same tuning but at different $\varepsilon$ as indicate in (a). (f) Rabi oscillations, taken at a different device tuning, demonstrating over 100 coherent $X_{\pi/2}$ rotations within a Rabi decay time. (g) Rabi oscillations, taken at a different device tuning, demonstrating a Rabi decay time longer than 1 $\mu$s. (h) $\Gamma_{Rabi}$ obtained by fitting to an
exponential decay showing a quadratically dependence on $dE_Q/d\varepsilon$. The different tunings are labeled with different colors (red, green, and blue). (i) Rabi decay rate showing a linear dependence on the microwave amplitude $A_{\varepsilon}$. Reprinted with permission from \cite{Thorgrimsson-2017} Copyright 2017, Springer Nature.}\label{3_Thorgrimsson}
\end{figure}

Following this path, it is of fundamental interest the detrimental effect of the environment and conversely how it is possible to increase significantly $T_2^\ast$ and $\Gamma_{Rabi}$ by appropriate tuning the qubit parameters and the operating points. In Ref. \cite{Thorgrimsson-2017} a very promising $T_2^\ast$ of 177 ns and a Rabi decay time $1/\Gamma_{Rabi}$ exceeding 1 $\mu$s are found, as from Fig. \ref{3_Thorgrimsson}.

 In particular Fig. \ref{3_Thorgrimsson}a shows both the energy of the qubit (black line) and its derivative with respect to the detuning (red line) as a function of detuning, showing the decrease in the slope $dE_Q/d\varepsilon$ with increasing $\varepsilon$. Figure \ref{3_Thorgrimsson}b–e shows Rabi oscillation measurements, performed by applying a microwave burst of duration $t_{RF}$ and acquired at the detunings labeled b–e in Fig. \ref{3_Thorgrimsson}a. In \ref{3_Thorgrimsson}f-g Rabi oscillations, taken at a different device tuning, demonstrate over 100 coherent X $\pi/2$ rotations within a Rabi decay time (f) and a Rabi decay time longer than 1 $\mu$s (g). Finally in \ref{3_Thorgrimsson}h-i $\Gamma_{Rabi}$ is obtained by fitting to an exponential decay and plotted as a function of $dE_Q/d\varepsilon$ (h) highlighting a quadratically dependence and as a function of the microwave amplitude $A_\varepsilon$ (i) showing a linear dependence. The different tunings are labeled with different colored dot.

A comparison between theory and experiment is reported in \cite{Abadillo-2018} where, by employing tight-binding simulations, they identify potential disorder profiles that induce behavior consistent with the experiments. Sweet spots where the decoherence caused by charge noise is suppressed are identified. How the interfacial atomic structure can be used in particular cases as a tool to enhance the fidelity of Si double-dot qubits has been explored. 

To summarize, Table \ref{collection} contains a collection of the main hybrid qubit experimental results extracted from different experimental papers.
\begin{center} 
\begin{table}[htbp!]
\begin{tabular}{ |p{2.8cm}|p{2cm}|p{2cm}|p{2cm}|p{2cm}|}
\hline
 $T_1$ & $T_2^{\ast}$ & $1/\Gamma_{Rabi}$ & X Gate Fidelity & Z Gate Fidelity \\
\hline
 $141\pm 12$ (ms) \cite{Shi-2012}& 177 (ns) \cite{Thorgrimsson-2017} & $\sim$ 1 ($\mu$s) \cite{Thorgrimsson-2017} &  93\% \cite{Kim-2015} & 96\% \cite{Kim-2015}\\ 
 \hline
\end{tabular}
\caption{Hybrid qubit experimental results collection.}\label{collection}
\end{table}
\end{center}

\section{Toward large scale fabrication of spin qubits in silicon}
In principle, the key advantages of the silicon platform consist of its capability to co-integrate both the physical substrate of the qubits and the control electronics within a single technology, and the fabrication of millions of physical qubits by a single process on a small chip. The existing silicon supply chain enables fabrication at the 7 nm technology node, while the 5 nm technology node is being developed. The main issues to fight are variability among qubits, as tiny fabrication differences at the level of single atom positioning are potentially harmful, and source of decoherence. As said, the latter may be mitigated by using isotopically pure silicon.

\subsection{From arrays of quantum dots to 300 mm wafers}

A two-dimensional arrangement of single spin qubits has been designed for instance in Ref.  \cite{Veldhorst-2017}. The proposal consists of a single spin qubits quantum computer based on CMOS technology organized according a surface code architecture.  

Another two-dimensional array crossbar architecture is reported in Ref. \cite{Li-2018}. It has a three-layer design to define qubit and tunnel barrier gates. It is based on shared control and a scalable number of control lines. The scheme is proposed for a two-dimensional array of quantum dots that can operate a large number of qubits with high fidelity, to support universal fault-tolerant quantum computation. The ability to shuttle qubits over large distances in principle provides means to realize quantum error correction schemes and quantum circuit implementations otherwise reserved to non-planar architectures.

The challenging problem of the increase of the heat load because of scaling versus the fixed amount of cooling power of dilution refrigerators is investigated in Ref. \cite{Dzurak-2019}. It consists of the first report of operation of two qubits confined by quantum dots at 1.5 K. In $^{28}$Si the single-qubit gate fidelity is 98.6\% and the coherence time $T_2^\ast= 2 \mu s$. The quantum dots are isolated from the electron reservoir. Coherent control of the qubits requires electrically-driven spin resonance (EDSR). 

Quantum dot based spin qubits are developed in the prospect of achieving high densities, all-electrical operation,
and integration onto an industrial platform. A system composed by two qubits has been demonstrated to overcome qubit crosstalk, state leakage and calibration issues \cite{watson2018programmable}. Such programmable two-qubits quantum processor has been shown to perform both the Deutsch-Josza and the Grover search algorithms at minimal resource scale. The fidelity characterized through quantum state tomography of Bell states is of about 85-89\%.
Such technology based on quantum dots, being is either Si-MOS or Si/SiGe, requires industrial class fabrication on 300 mm wafers.
To achieve such goal, the footprint of the actual qubit which includes the wiring according to the design rules of a technology node has been evaluated \cite{Rotta-2017}. The choice of all-electrical hybrid spin qubits requires spins to be sufficiently close to allow an operation time compatible with fault tolerant quantum computing. Such constraint limits the candidate technology nodes below 14 nm node while time operation windows span the range 1-10 GHz. 
Intel is developing a 300mm process line for spin qubit devices using immersion lithography and isotopically pure epitaxial silicon layers \cite{watson2018programmable}. Transistors and quantum dot devices are co-fabricated on the same wafer. Such technology requires high quality material deposition, including highly purified silicon, and high process control \cite{clarke2019spin}.

\subsection{Packing qubits: comparison with other qubit technologies}

The time past between the first demonstration of a two-qubits algorithm by superconductor qubits \cite{dicarlo2009demonstration} and the first gate model programmable chip generation with several qubits is less than a decade. Spin qubits in silicon and a two-qubits quantum algorithm have been demonstrated up to now with a delay of about 7-8 years so if such pattern is respected, after the first two-qubits quantum algorithm implementation in silicon of 2018 \cite{watson2018programmable}, the first programmable gate-model silicon quantum chip could be fabricated by 2025.   


The long term aim is to achieve a number of physical qubits of the order of  $10^6-10^8$ so to run generic quantum algorithms. Moving toward large scale quantum chips raises issues related to the physical footprint of all the control connections from the qubits to the outside world, the cross-talk minimisation and the heat dissipation.    

In Table \ref{comparison} the footprint of the physical qubits by different technologies is compared. The density of quantum information, i.e. the number of physical qubits per unit area, is reported for the silicon qubit (single-spin and hybrid qubit) in comparison with the qubits of existing quantum computers namely  superconductive and trapped ions qubits. 
\begin{center} 
\begin{table}[h]
\begin{tabular}{ |p{1.8cm}|p{2.1cm}|p{2.1cm}|p{2.1cm}|p{2.2cm}|p{2.2cm}|p{1.3cm}| }
\hline
 & Semiconductor Single-Spin qubit & Semiconductor Hybrid qubit (Steane code)& Semiconductor Hybrid qubit (Surface code)& Superconductor Flux qubit (DWave like)& Superconductor Transmon qubit (IBM like)  & Trapped Ion qubit\\
\hline
$Mqb_{ph}/cm^2$ & 8000 & 830 & $100\times 10^{2}$ & $8\times 10^{-4}$ & $10^{-5}$  & $2\times 10^{-5}$\\ 
\hline
$A_{chip}(mm^2)$& 25 & 240 & 20 & $25\times 10^{7}$ & $2\times 10^{10}$ & $10^{10}$\\
\hline
Reference & \cite{Li-2018} & \cite{Rotta-2017}& \cite{Rotta-2017}& \cite{Harris-2010} & \cite{Gambetta-2017,Kandala-2017}  & \cite{Lekitsch-2017}\\ 
\hline
\end{tabular}
\caption{Number of physical qubits per unit surface and area covered by 2 billions of physical qubits. The silicon hybrid qubit footprint refers to the 7 nm technology node.}\label{comparison}
\end{table}
\end{center}

The footprint of silicon qubits is significantly smaller than superconductors and trapped ions, which may raise concerns when scaling above hundreds of thousands.

\subsection{Cryogenic control electronics}

The issue of the control electronics is discussed in Ref.  \cite{Schaal-2019} where an hybrid solution between conventional and quantum electronics is suggested. A circuit operating at near-absolute-zero temperature divided into cells is detailed, each cell containing a control field-effect transistor and a quantum dot device, formed in the channel of a nanowire transistor. The readout is done by interfacing the cells with a single radio-frequency resonator. Both single charge sensing for spin to charge conversion and fast rf-reflectometry readout scheme can be used. The idea beyond this approach is to reduce the number of input lines per qubit when going to large-scale device arrays.

The scalability of the hybrid qubit is the subject of the work done in \cite{Rotta-2016} and in \cite{Rotta-2017} where the footprint of silicon spin qubits is evaluated according to industrial fabrication constraints. The operation bandwidth of such qubits at the technology node below 14 nm is compatible with ordinary 1-10 GHz control operations which in turn are allowed by standard coaxaial lines. 

The cryogenic Field-Programmable Gate Arrays (FPGA) \cite{Homulle-2017} and electronics in general have been intensively discussed by several groups \cite{hornibrook2015cryogenic,patra2017cryo}. Multiplexing and the hope of operating spin qubits at 1.5 K \cite{Dzurak-2019} significantly relaxes the heavy requirements due to the thermal load carried by several qubits. 

\section{Conclusions and perspectives}
It is now universally accepted that a full-scale quantum processor would have applications in a variety of different scientific, social and economical contexts from finance to the security and medical sectors. Quantum computers may play a special role by their combination with artificial intelligence  towards quantum machine learning \cite{prati2017quantum}.
Despite a delay of about 7-8 years in terms of maturity with respect for instance to superconductor qubits, silicon is definitely in strong competition with the two technologies used to fabricate commercial quantum computers, namely superconductors and trapped ions by corporate giants (Google, Rigetti, IBM, Alibaba and IonQ). 

The fabrication of a complete silicon fault-tolerant architecture is an ambitious task, but the possibility of using the well assessed semiconductor manufacturing paves the way towards the large-scale quantum computation era based on the same silicon chip technology at the heart of our current information age. Moreover silicon quantum computer lies on a technology platform ideal for scaling up to the large numbers of qubits needed for universal quantum computing. 


If on one side the CMOS technology for the qubit fabrication and the integration with the cryogenic control represent ambitious and challenging tasks, the recent advancements in this field from theory and experiments that we have highlighted in this perspective article demonstrate that they are not unrealistic. The silicon route must be pursued and the goal could be achieved together with the enhancement of the fabrication techniques especially in $^{28}$Si. In conclusion there is a serious chance that silicon will provide a feasible technological platform to support future universal quantum information processing.
\bibliographystyle{unsrt}
\bibliography{Ref}
\end{document}